\documentclass[aps,amsmath,amssymb,floatfix,twocolumn,prb, longbibliography]{revtex4-1}
\usepackage{graphicx,amsfonts,bm,amsmath,verbatim,color,array}
\usepackage{epstopdf}

\begin{document}
\title{Suppression of Hall number due to charge density wave order in high-$T_c$ cuprates}

\author{Gargee Sharma$^1$}
\author{S. Nandy$^{2}$}
\author{A. Taraphder$^{2,3}$}
\author{Sumanta Tewari$^{2,4}$}

\affiliation{$^1$Department of Physics, Virginia Tech, Blacksburg, VA 24061, U.S.A\\
$^{2}$ Department of Physics, Indian Institute of Technology Kharagpur, W.B. 721302, India\\
$^{3}$ Centre for Theoretical Studies, Indian Institute of Technology Kharagpur, W.B. 721302, India\\
$^4$Department of Physics and Astronomy, Clemson University, Clemson, SC 29634,U.S.A}

\begin{abstract}
Understanding the pseudogap phase in hole-doped high temperature cuprate superconductors remains a central challenge in condensed matter physics. From a host of recent experiments there is now compelling evidence of translational symmetry breaking charge density wave (CDW) order  in a wide range of doping inside this phase. Two distinct types of incommensurate charge order -- bidirectional at zero or low magnetic fields and unidirectional at high magnetic fields close to the upper critical field $H_{c2}$ -- have been reported so far in approximately the same doping range between $p\simeq 0.08$ and $p\simeq 0.16$. In concurrent developments, recent high field Hall experiments have also revealed two indirect but striking signatures of Fermi surface reconstruction in the pseudogap phase, namely, a sign change of the Hall coefficient to negative values at low temperatures at intermediate range of hole doping and a rapid suppression of the positive Hall number without change in sign near optimal doping $p \sim 0.19$. We show that the assumption of a unidirectional incommensurate CDW (with or without a coexisting weak bidirectional order) at high magnetic fields near optimal doping and a coexistence of both types of orders of approximately equal magnitude at high magnetic fields at intermediate range of doping may help explain the striking behavior of low temperature Hall effect in the entire pseudogap phase.
\end{abstract}

\maketitle

\section{Introduction}
The parent compounds of high-$T_c$ copper oxide superconductors at half-filling (hole doping $p \simeq 0$) are antiferromagnetic Mott insulators. At high hole doping $p \gtrsim 0.2$ the electrons form a fairly conventional metallic state with the Fermi surface given by a large hole-like cylinder with a carrier density $n_c \simeq 1+p$. The pseudogap phase in the intermediate range of hole doping which connects the antiferromagnetic Mott insulator at low $p$ with the metallic state at large $p$ remains a central puzzle in the physics of hole-doped cuprate superconductors \cite{Norman,Taillefer}.
In attempts to demystify the physics in the pseudogap regime, recent measurements of high field ($\sim H_{c2}$) and low temperature Hall number in YBCO and other cuprate superconductors as a function of hole doping have revealed signatures of a Fermi surface reconstruction near optimal doping $\sim p^*\sim 0.19$ \cite{Bala1,Bala2,Bala3,Badoux:2016,Laliberte:2016,Collignon:2016}.
Badoux et al. \cite{Badoux:2016} has recently reported a rapid drop of high field and low temperature positive Hall number
with decreasing doping via a quantum phase transition (QPT) or a sharp crossover near optimal doping $p=p^*\sim 0.19$.  Since the zero-temperature Hall number $n_H =1/R_H = \sigma_{xx}\sigma_{yy}/\sigma_{xy}=n_ce/B$ where $R_H$ is the Hall constant and the last equality is for conventional Drude theory with $n_c$ the carrier density and $B$ the applied magnetic field, it
provides information about the volume
enclosed by the Fermi surface - i.e. is equivalent to the
electronic density per unit cell of the crystal. Therefore, a drastic drop in $n_H$ below optimal doping possibly indicates a drastic reconstruction of a large Fermi surface enclosing a volume corresponding
to a density $n_c \simeq 1 + p$ of holes at large doping,
to small pockets with a volume corresponding to
a hole-density $p$ in the underdoped regime.
Remarkably, at lower doping values for moderately underdoped cuprates, the high-field, low-temperature, Hall and Seebeck coefficients turn negative, indicating a second reconstruction of the Fermi surface from hole to electron pockets \cite{LeBoeuf:2007,Chang:2010,Laliberte:2011}, also consistent with quantum oscillation experiments \cite{Leyraud}. The story of the Fermi surface in pseudogap phase of the cuprates is thus complex, possibly marked by two successive reconstructions, one from a large hole-like surface to small hole pockets near optimal doping and a second one from hole- to electron-like pockets at intermediate doping. The primary focus of this paper is the rapid suppression (while staying positive) of the high-field low-temperature Hall number near optimal doping, possibly indicating a Fermi surface reconstruction from large hole-like Fermi surface to small hole pockets below optimal doping, in terms of an incommensurate uni-directional charge density wave state which has been revealed in recent high-field X-ray and NMR experiments. We also briefly include a discussion of the behavior of the Hall coefficient in the moderately underdoped regime~\cite{Seo:2014}, where the Hall coefficient has been observed to change sign at low temperature, for the sake of completeness.

The presence of charge density wave (CDW) correlations in the pseudogap phase of copper oxide superconductors is now well established \cite{Damascelli,Wu:2015,Wu:2011,Wu:2013,Chang:2012,Ghiringhelli:2012,Comin,Neto,Tabis,Croft,Gerber,Chang,Jang,Blackburn,Blanco,Lali}. In zero or low magnetic fields the CDW order has been observed in multiple x-ray scattering experiments \cite{Damascelli,Chang:2012,Ghiringhelli:2012,Comin,Neto,Tabis,Croft} as well as nuclear magnetic resonance (NMR) \cite{Wu:2015}, and is static, short range correlated, and bidirectional (i.e., charge density is modulated in both Cu-O bond directions in the CuO$_2$ planes). In YBa$_2$Cu$_3$O$_x$ (YBCO) a second charge density wave order has also been found in high magnetic fields and lower temperatures \cite{Wu:2015,Wu:2011,Wu:2013,Gerber,Chang,Jang,Lali} and is long range ordered, but essentially unidirectional. So far, this high field CDW has been observed in NMR \cite{Wu:2015,Wu:2011,Wu:2013}, and more recently, x-ray scattering experiments in high magnetic fields \cite{Gerber,Chang,Jang}. In contrast to the low-field bidirectional short-range-correlated CDW with an onset temperature close to $150 K$, the high-field unidirectional long-range-correlated CDW onsets above a critical magnetic field proportional to the resistive upper critical field $H_{c2}$ with heavily suppressed superconductivity
and at a temperature close to the superconducting transition temperature $T_c$ at zero magnetic field. Since the observation of the unidirectional CDW requires x-ray scattering at high magnetic fields, which is challenging because the scattering signals are weak, to date the direct experimental observation of this high-field phase has remained limited, only around the moderate doping levels $p \sim 0.12$. In more recent high magnetic field sound velocity measurements \cite{Lali} it has been found that the high field unidirectional charge order exists in more or less the same doping range $0.08\lesssim p \lesssim 0.16$ as its low-field bidirectional counterpart, although the intrinsic connection between the two distinct charge orders, if any, is still unclear. Nonetheless, because of its large correlation volume and sharp onset in magnetic field and temperature, both in contrast to the low-field short-range-correlated CDW at higher temperatures, it has been argued that the ground state competing order in clean superconducting YBCO is a long-range-ordered, incommensurate, CDW in which charge modulation is unidirectional \cite{Jang}.
The primary focus of our current work is an explanation of the rapid suppression of the Hall number near optimal doping (while staying positive) indicating a Fermi surface reconstruction in terms of uni-directional CDW which has been observed in multiple CDW experiments. We also add a discussion of underdoped in terms of bi-directional CDW~\cite{Seo:2014} for the sake of completeness. 

 On the theoretical side, the Fermi surface reconstruction near optimal doping from large hole-like Fermi surface at higher doping to small hole pockets at lower doping has been attributed to a quantum phase transition to an assumed long range ordered $d$-density wave \cite{Tewari} or antiferromagnetic \cite{Eberlein} phase or an assumed nematic transition \cite{Maharaj} near optimal doping. However, experimental evidence for translational symmetry breaking bond currents or long ranged antiferromagnetic order has so far not been found in the relevant range of doping and magnetic fields.

In this work, we assume that the Fermi surface reconstruction at high magnetic fields ($H\gtrsim H_{c2}$) near optimal doping is caused by a  high-field long-range-ordered state, namely, incommensurate unidirectional CDW, that has been observed unambiguously in the pseudogap phase of the cuprates, although not near optimal doping. A similar long range ordered unidirectional CDW with $C_4$ symmetry breaking was considered earlier to explain the high field quantum oscillation experiments in the pseudogap phase \cite{Yao}.
 We find that the reconstructed Fermi surface in this state consists of hole pockets, in contrast to the Fermi surface of the low-field bidirectional CDW which is dominated by electron pockets. This is consistent with earlier calculations on bidirectional CDW which first proposed the existence of electron pocket \cite{Seb1} in this state and also on unidirectional CDW \cite{Seb2} which showed the existence of hole pockets.
 Starting from the Kubo formula which reduces to the semi-classical Boltzmann equations in appropriate limits, we calculate the  Hall number in the unidirectional CDW state as a function of  hole  doping  below optimal doping. We find that the assumed onset of this CDW (with or without a weak bidirectional component) explains the rapid drop in Hall number near optimal doping as observed in recent experiments~\cite{Badoux:2016, Laliberte:2016, Collignon:2016}. 
 Adding a bidirectional component of equal magnitude to the unidirectional order parameter at intermediate range of hole doping, which is now supported by experiments \cite{Damascelli,Wu:2015,Wu:2011,Wu:2013,Chang:2012,Ghiringhelli:2012,Comin,Neto,Tabis,Croft,Gerber,Chang,Jang,Blackburn,Blanco,Lali,Wu:2011,Wu:2013,Gerber,Chang,Jang,Lali}, has the effect of including an electron pocket in the Fermi surface. The corresponding Hamiltonian reduces the zero temperature Hall number to negative values. Note that, recent high field x-ray scattering \cite{Gerber,Chang,Jang} and sound velocity \cite{Lali} measurements point to the co-existence of unidirectional and bidirectional CDW in the moderate range of hole doping, but not close to optimal doping. Our calculations explain the salient features of the high-field Hall effect experiments in the pseudogap phase, namely, sharp drop below optimal doping and negative values at moderate underdoping,
 entirely in terms of coexisting unidirectional and bidirectional charge orders (with weak bidirectional component near optimal doping), both of which have been unambiguously observed in the pseudogap phase of the cuprates at relevant scales of magnetic field and temperature.
  
The rapid suppression of the positive Hall number without a change in sign near optimal doping $p\sim 0.19$ due to Fermi surface reconstruction remains a pressing unresolved issue in cuprates.  In the light of recent experiments~\cite{Bala1,Bala2,Bala3,Badoux:2016,Laliberte:2016,Collignon:2016}, our results are  important as the assumption of a uni-directional CDW can very well explain the salient experimental features near optimal doping.
A series of recent papers attempt to explain the Fermi surface reconstruction and the drop in the Hall number using other ordered states~\cite{Maharaj,Eberlein,Tewari}, none of which have been confirmed experimentally. This makes our proposal highly relevant due to ample evidence of the CDW phase in cuprates~\cite{Damascelli,Wu:2015,Wu:2011,Wu:2013,Chang:2012,Ghiringhelli:2012,Comin,Neto,Tabis,Croft,Gerber,Chang,Jang,Blackburn,Blanco,Lali}.  
  
This paper is organized as follows: In Section II we introduce the mean field Hamiltonian for unidirectional CDW state and calculate its quasiparticle Fermi surface. In Section III we calculate the Hall number from the unidirectional CDW state which exhibits a drastic drop below the hole doping $p^*$. In Section IV we calculate the quasiparticle spectrum, Fermi surface, and the Hall resistance from coexisting unidirectional and bidirectional CDW order parameters. We discuss our results in Section V and conclude in Section VI. 

\section{Hamiltonian and the Fermi surface}
We begin by writing down the real space mean-field Hamiltonian for the unidirectional CDW state
on a two-dimensional square lattice given by,
\begin{eqnarray}
H_{CDW}^{uni} &= \sum\limits_{\mathbf{r},\mathbf{a},\sigma} [W_{\mathbf{a}} e^{i\mathbf{Q}\cdot(\mathbf{r}+\mathbf{a}/2)} c^{\dagger}_{\mathbf{r}+\mathbf{a},\sigma}c_{\mathbf{r},\sigma}\nonumber  + h.c],\\
\label{BDW1}
\end{eqnarray}
where in the sum $\mathbf{r}$ denotes the lattice sites, and the vector $\mathbf{a}$ represents all the nearest neighbors vectors. The operator $c_{\mathbf{r},\sigma}$ annihilates an electron of spin $\sigma$ at site $\mathbf{r}$. We assume the order parameter $W_{\mathbf{a}}$ with a $d$-wave-like form factor $W_{\pm\hat{x}} = -W_{\pm\hat{y}} = W_0/2$, where $W_0$ is the bare magnitude of the order parameter~\cite{Andrea:2014}. When  $\mathbf{Q}= (1/N,0)r.l.u$, the above equation represents a commensurate CDW with a periodicity of $N$ lattice vectors. Experimental evidence suggests a slight incommensuration in the scattering vector $\mathbf{Q}$. For our calculations we assume a scattering vector $\mathbf{Q}=(q_0+\delta,0)$, where $q_0=1/3$, and $\delta$ is the small incommensuration with respect to the underlying lattice.
\begin{figure}
\includegraphics[scale=0.215]{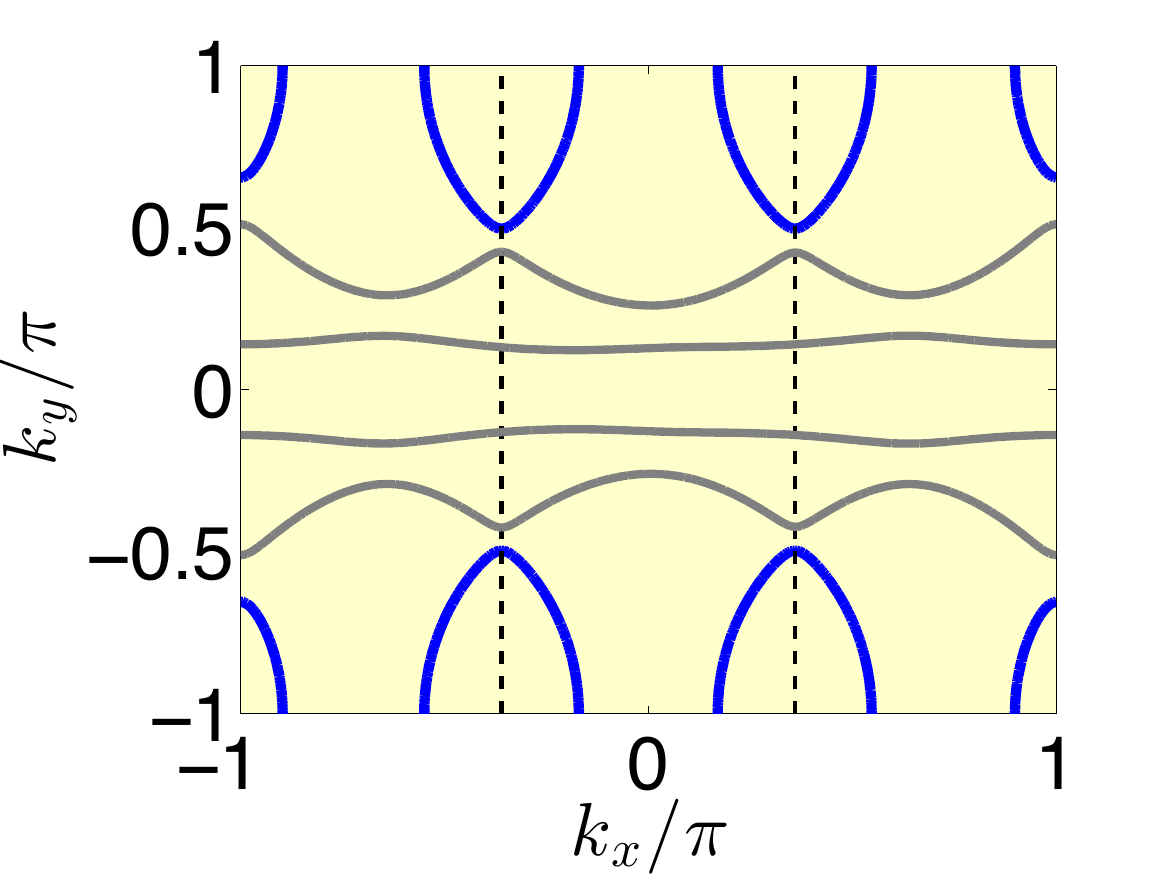}
\includegraphics[scale=0.215]{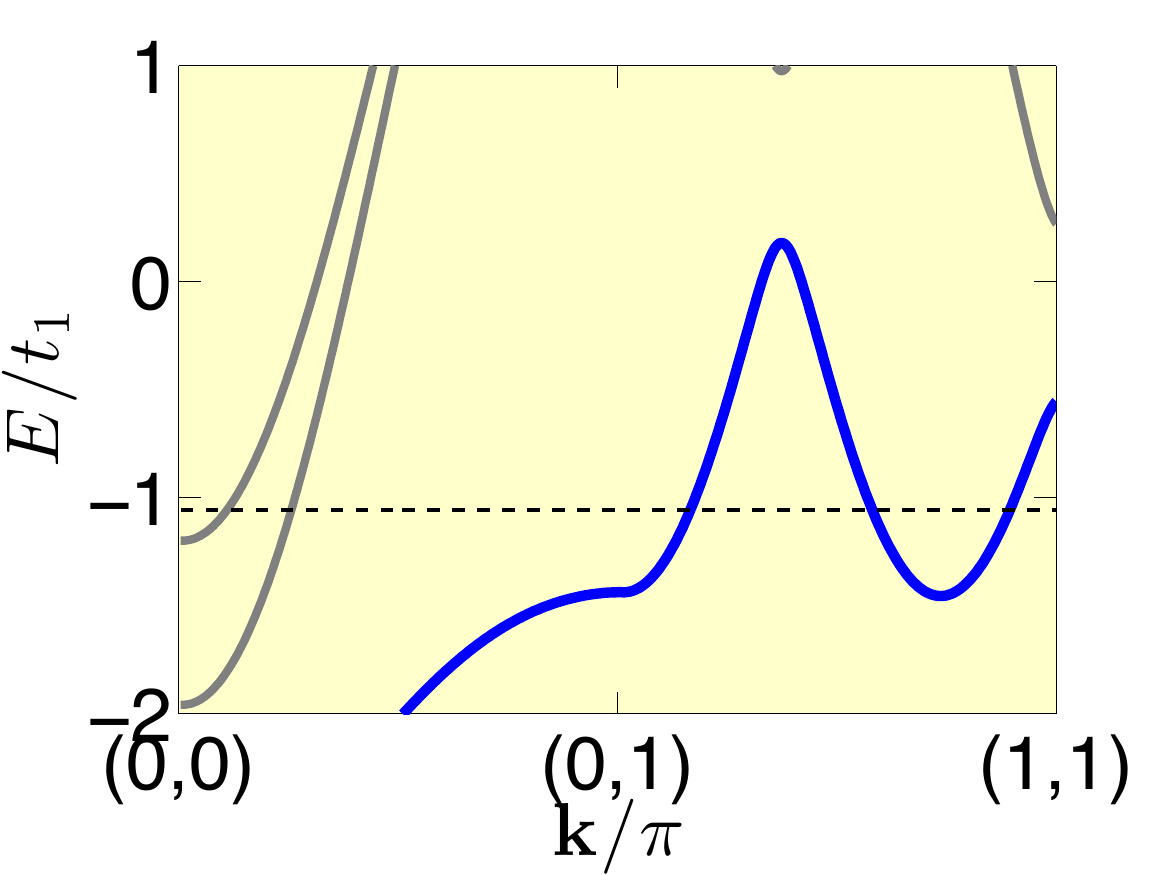}
\includegraphics[scale=0.215]{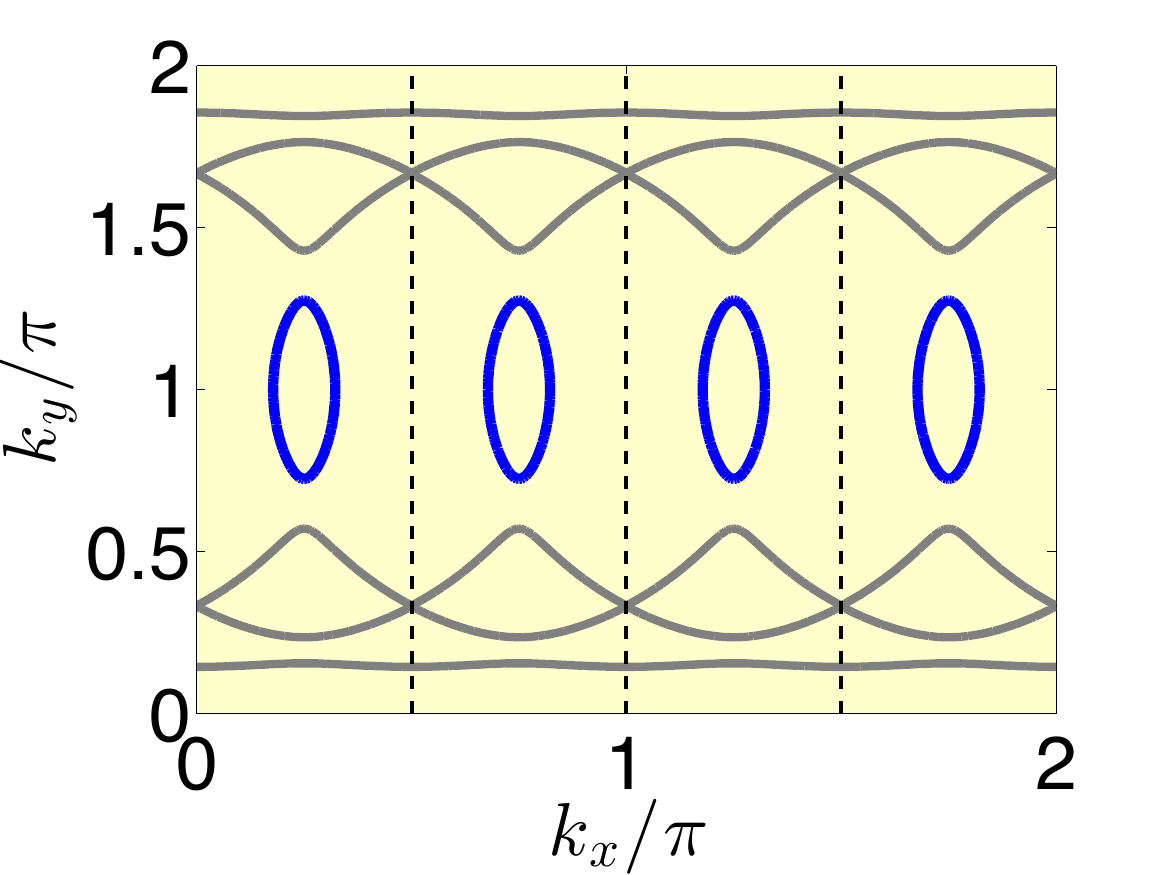}
\includegraphics[scale=0.215]{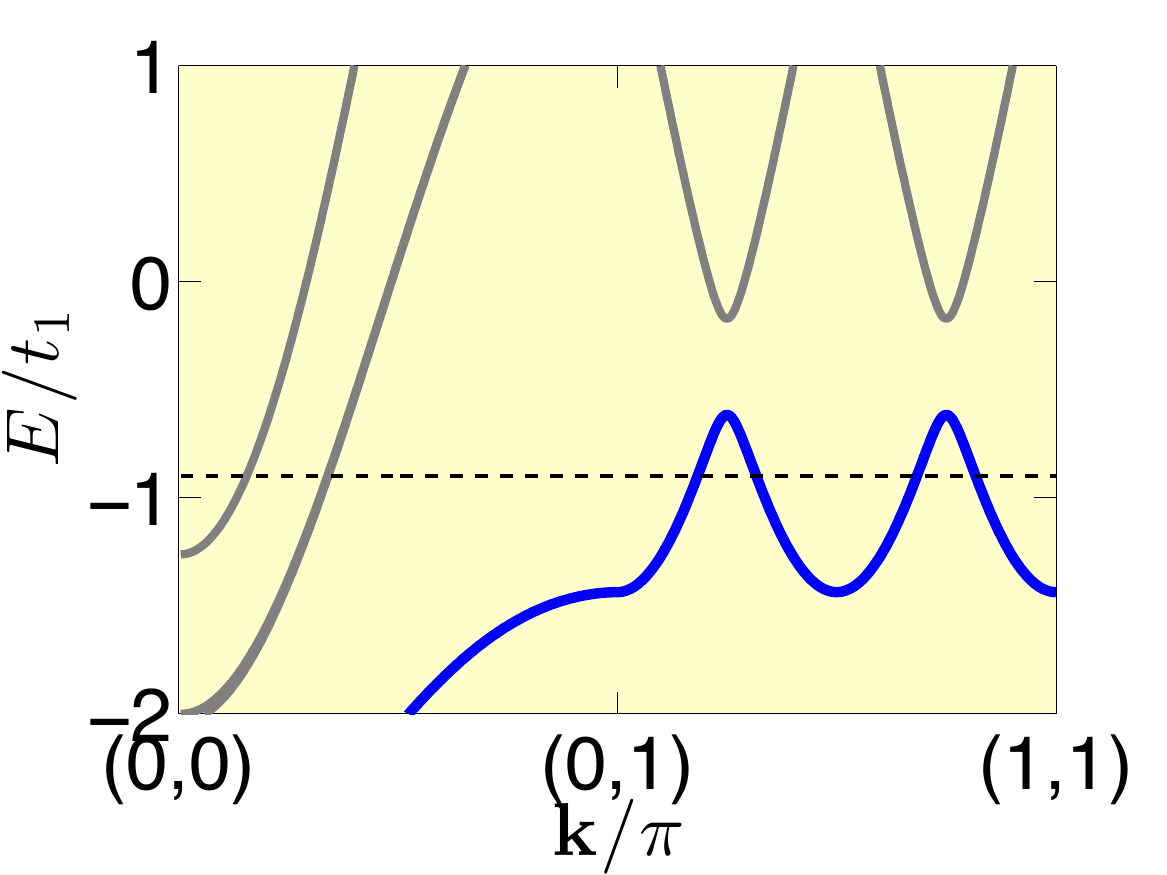}
\caption{\textit{Top panel:} (color online) Quasiparticle Fermi surface (left) plotted over the unfolded Brillouin zone, and the energy dispersion (right) for the uni-directional CDW state with scattering vector $\mathbf{Q}=(q=q_0+\delta,0)r.l.u$ (with a slight incommensuration of $\delta=0.03$ from $q_0=1/3$) for doping $p=0.16$, and $W_0=0.2t_1$. The hole pockets centered at $(\pm q\pi,\pm\pi)$ are displayed in blue. The RBZ is enclosed within the two dashed lines on the left plot. The dashed line on the right plot indicates the chemical potential. \textit{Bottom panel:} Reconstructed Fermi surface (left) and the energy dispersion (right) for a unidirectional CDW order parameter, assuming a four-component reconstruction of the Brillouin zone with the scattering vector $\mathbf{Q}=(q,0)r.l.u$ where $q=0.25$. The hole-pockets are displayed in blue.}
\label{Fig_FS_ARPES_1}
\end{figure}
\begin{figure}
	\includegraphics[scale=0.43]{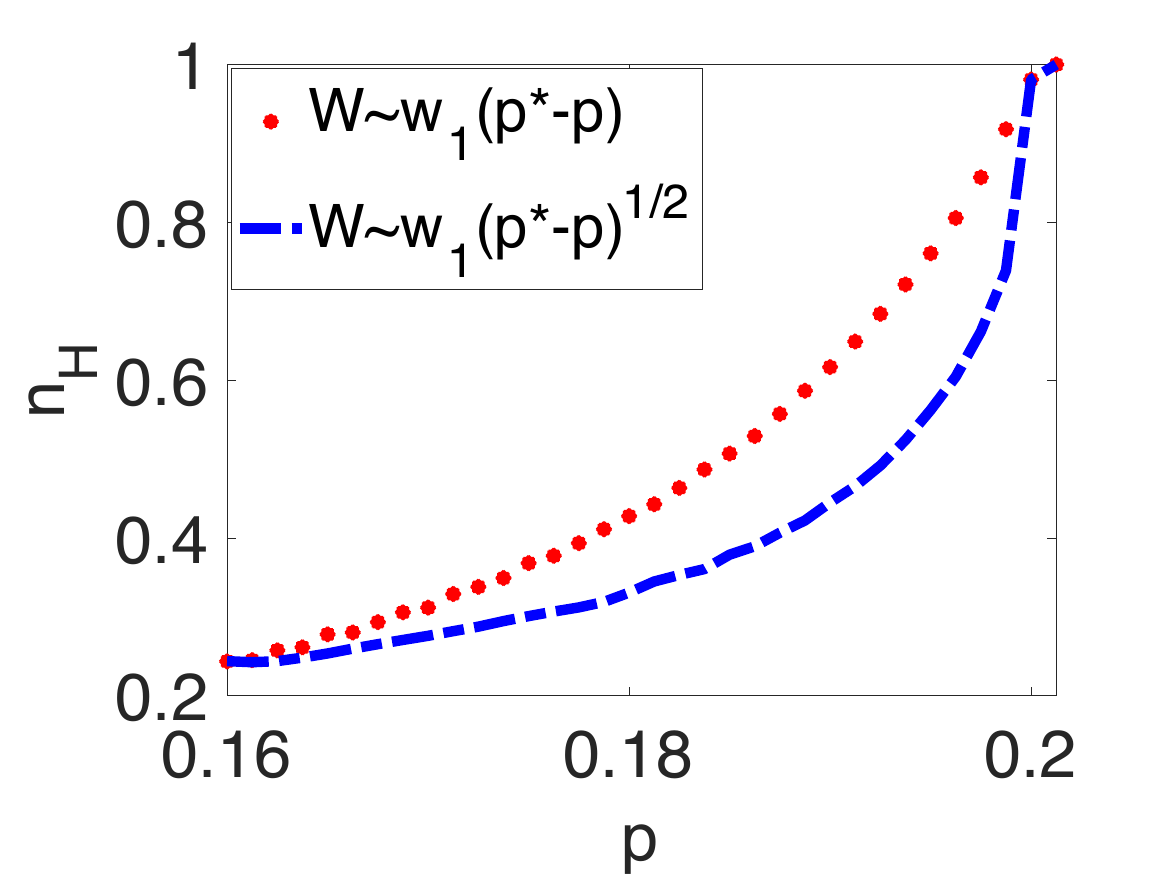}
	\caption{(color online) Hall number $n_H$ (normalized w.r.t the Hall number at $p^*=0.20$) from the unidirectional CDW state plotted between hole-doping values $p=0.16$ and $p=0.2$. Above $p=p^{*}=0.20$, the CDW order parameter is zero. The parameter $w_1$ is chosen such that the bare CDW order parameter $W$ at $p=0.16$ is $W=0.34t_1$. Results with different phenomenological dependence of the CDW order parameter $W$ on hole-doping $p$ (linear and square-root) are shown. We note the rapid drop in the Hall number below $p^{*}$ to about 25\% at $p=0.16$. The parameters chosen were $t_1=0.3eV$ and $t_2=0.21t_1$. }
	\label{Fig_Hall_number}
\end{figure}
\begin{figure}
	\includegraphics[scale=0.35]{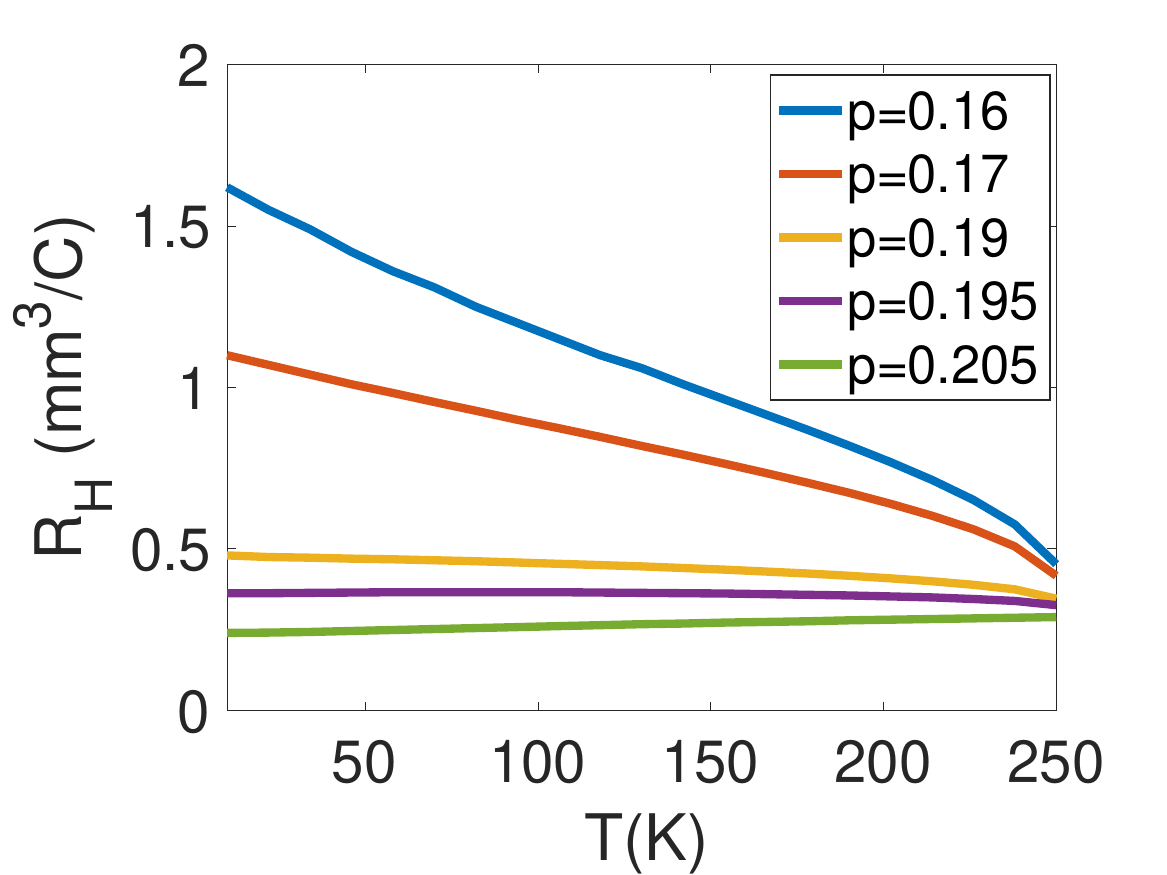}
	\includegraphics[scale=0.35]{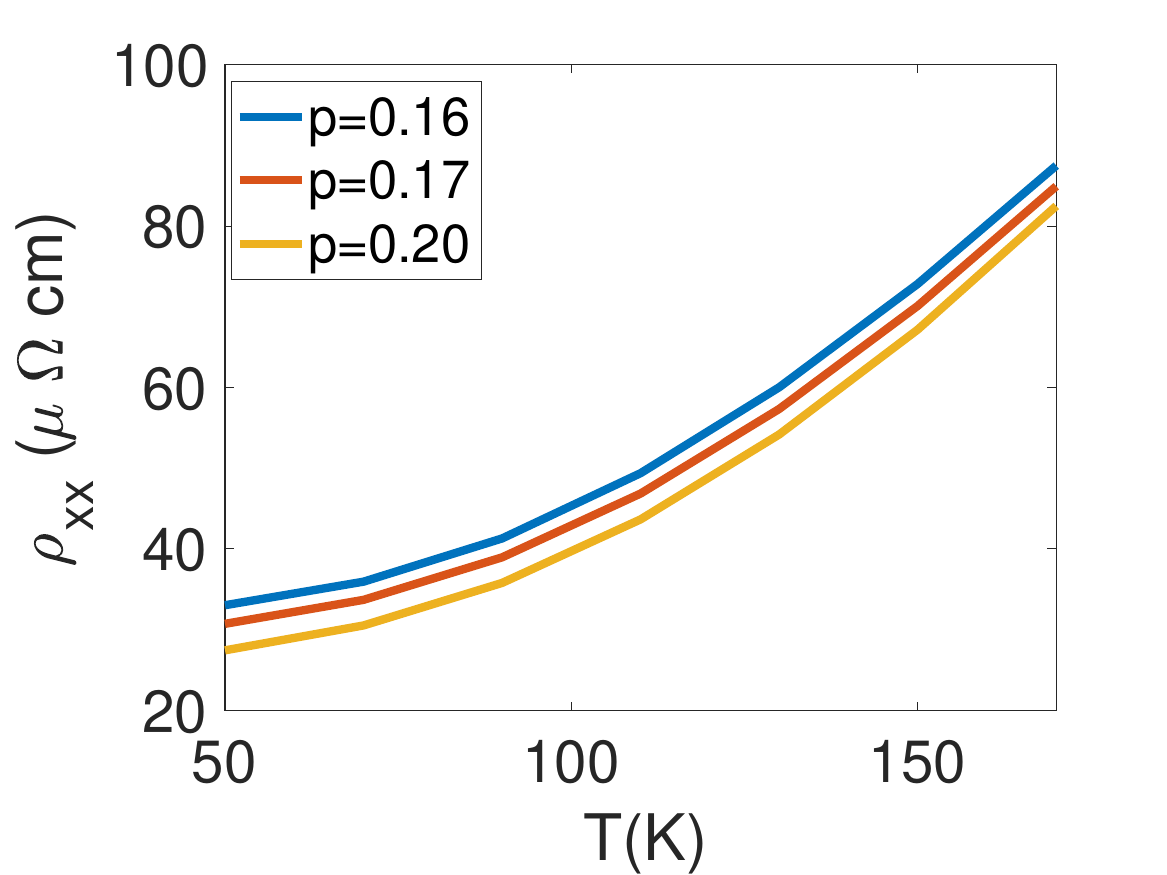}
	\includegraphics[scale=0.35]{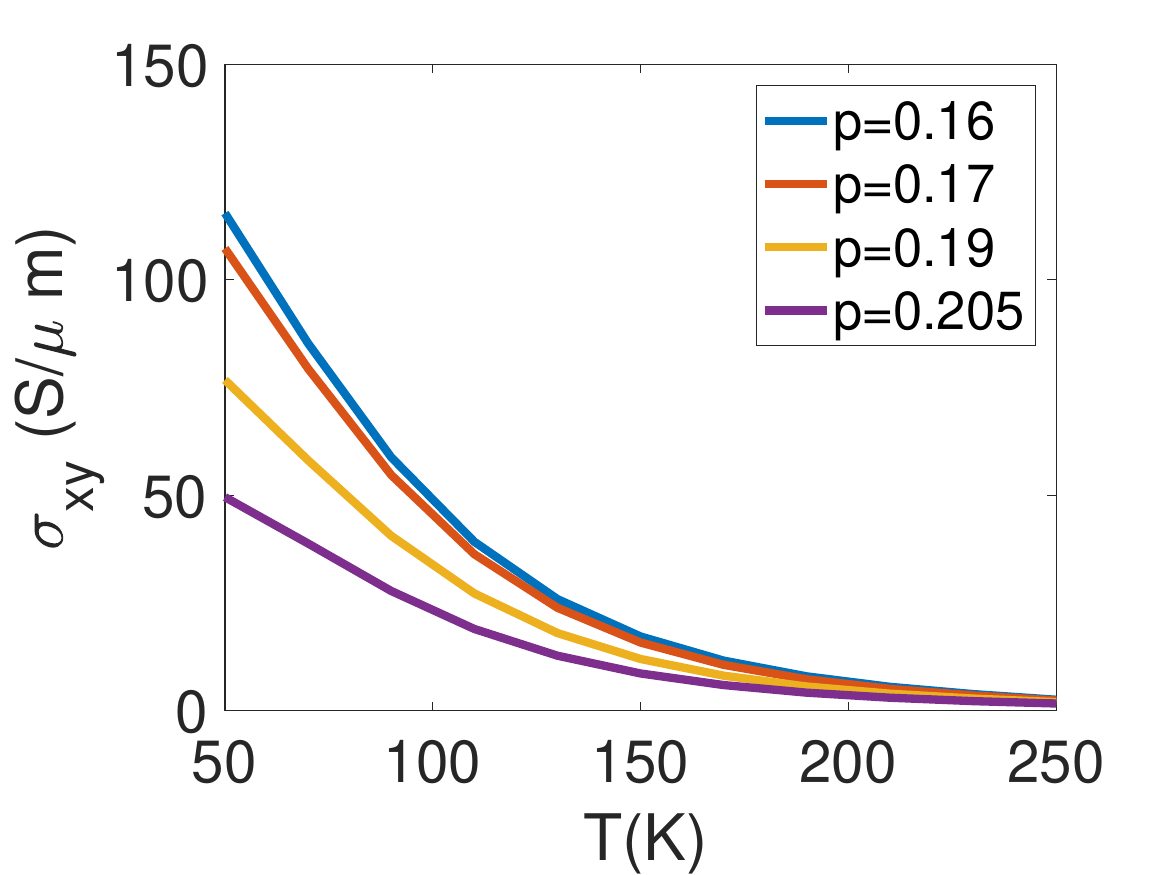}
	\caption{(color online) Hall resistance ($R_H$), longitudinal resistivity ($\rho_{xx}$), and Hall conductivity ($\sigma_{xy}$) as a function of temperature for various doping values in the actual experimental units, as predicted by a uni-directional CDW order. Here we use a standard mean field dependence of the CDW order parameter ($W\sim\sqrt{1-T/T_{CDW}})$. The order of magnitude of the above quantities, and the behavior w.r.t. temperature and doping match with the corresponding experimentally observed quantities in Ref~\onlinecite{Badoux:2016}.}
	\label{Fig_rh_rhoxx_sxy}
\end{figure}
The mean-field CDW Hamiltonian in momentum space is given by,
\begin{eqnarray}
H_{CDW}^{uni}&=W_0 \sum\limits_{\mathbf{k},\sigma}[ \left(\cos k_x -\cos k_y \right)c^{\dagger}_{\mathbf{k}+{{\mathbf{Q}}/{2}},\sigma}c_{\mathbf{k}-{\mathbf{Q}}/{2},\sigma}]\nonumber \\
&+ h.c,
\label{BDW2}
\end{eqnarray}
where $c_{\mathbf{k},\sigma}$ is the Fourier transform of $c_{\mathbf{r},\sigma}$. Combined with the 2D quasiparticle dispersion, the total Hamiltonian $H_{MF}$ for the system can be written as,
\begin{eqnarray}
H_{MF} = \sum\limits_{\mathbf{k},\sigma} \epsilon_{\mathbf{k}} c^{\dagger}_{\mathbf{k},\sigma}c_{\mathbf{k},\sigma} + H_{CDW}^{uni},
\label{H_MF_CDW1}
\end{eqnarray}
where
$\epsilon_{\mathbf{k}}=-2t_1(\cos k_xa + \cos k_ya ) + 4t_2 \cos k_xa \cos k_ya \nonumber - 2t_3 (\cos 2k_xa  + \cos 2k_ya )$
and $t_1$, $t_2$ and $t_3$ are the nearest-neighbor, next-nearest-neighbor, and next-to-next-neighbor hopping parameters, and $a$ is the lattice constant. Unless otherwise specified, for our calculations we choose the parameters $t_1=0.3$ $eV$, $t_2=0.3t_1$ and $t_3=0.1t_2$, consistent with earlier work on these systems~\cite{Chakravarty-ARPES}.
We assume that the interactions primarily give rise to a nonzero order parameter $W_0$ and thus ignore the residual interactions between the quasiparticles.
\begin{figure}
	\includegraphics[scale=0.35]{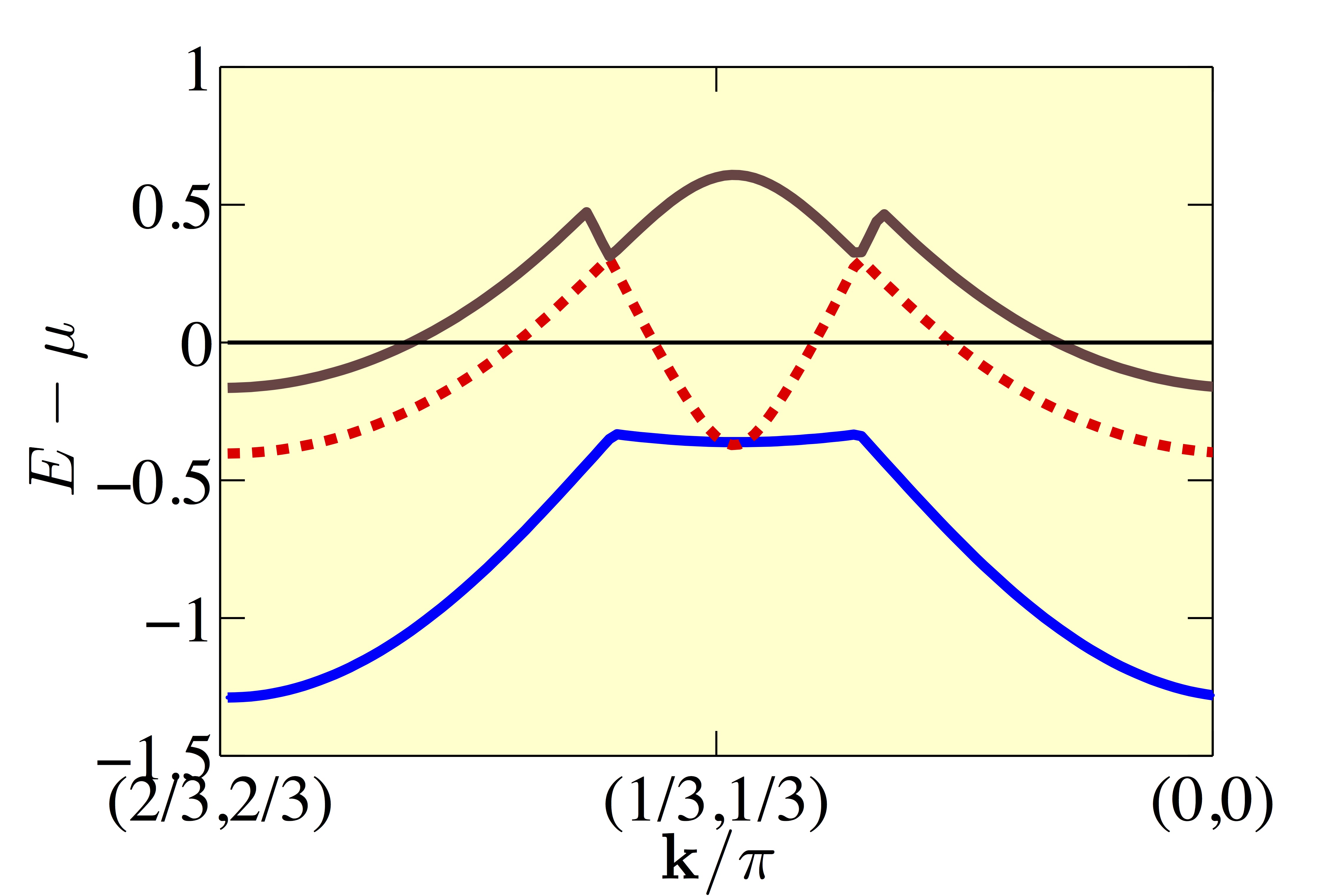}
	\includegraphics[scale=0.36]{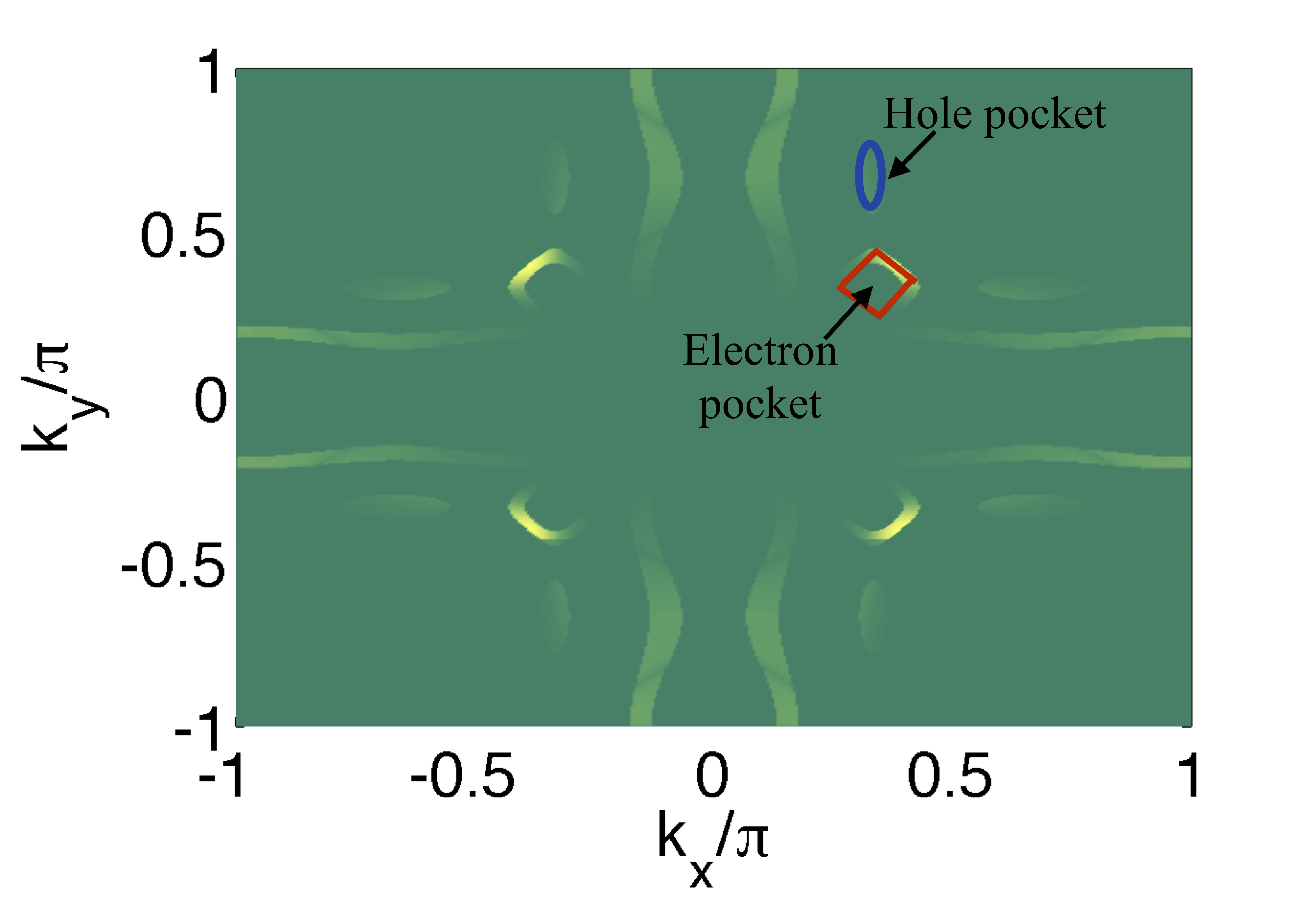}
	\caption{ (color online) Energy spectrum (top) near the Fermi energy and the corresponding ARPES spectral function $A(\omega=0,\mathbf{k})$ (bottom) from coexisting unidirectional and bidirectional CDW orders. The existence of electron pocket centered at $\mathbf{k}=(\pi/3,\pi/3)$  is specifically highlighted by a red rhombus in the plot for ARPES spectral function. The hole pocket at $\mathbf{k}=(\pi/3,2\pi/3)$ is also highlighted by a blue oval. The ratio between unidirectional and bidirectional CDW orders was taken to be 2:3, with a bare magnitude $W\sim 0.22t_1$.}
	\label{Fig_Coex_FS}
\end{figure}

The Hamiltonian in Eq.~\ref{H_MF_CDW1} can most easily be written by coupling wave vector $\textbf{k}$, confined to a properly defined reduced Brillouin zone (RBZ), with wave vectors translated by the CDW wave vector $\textbf{Q}$, i.e., $\textbf{k} \rightarrow \textbf{k}+ n_x Q_x \hat{x}+n_y Q_y \hat{y}$ where $n_x,n_y$ are integers denoting translations in the two-dimensional reciprocal space. Strictly speaking, for incommensurate systems this procedure results in an infinite dimensional Hamiltonian matrix and infinite number of bands. However, for the incommensuration $\delta \ll Q$, we can approximate the relevant energy eigenvalues by partitioning the unfolded BZ and defining energy bands over each BZ sector.  Using this approach, we write the Hamiltonian in terms of a three component operator $\Psi_{\mathbf{k},\sigma}$ as
\begin{equation}
H_{MF} = \sum_{\mathbf{k}\in RBZ, \sigma } {\Psi^{\dagger}_{\mathbf{k},\sigma} H(\mathbf{k}) \Psi_{\mathbf{k},\sigma}}
\label{HMF}
\end{equation}
where the sum is over the reduced Brillouin zone (RBZ) defined by ($0<|k_x|<\pi/3$, $0<|k_y|<\pi$), $\Psi^\dagger_{\mathbf{k}}=(c^\dagger_{\mathbf{k},\sigma},c^\dagger_{\mathbf{k}+\mathbf{Q},\sigma},c^\dagger_{\mathbf{k}-\mathbf{Q},\sigma})$ and  $H(\mathbf{k})$ is
\begin{equation}
H(\mathbf{k})=\left( \begin{array}{ccc}
\epsilon_{\mathbf{k}} & w_{12} & w_{13}  \\
w_{21} & \epsilon_{\mathbf{k}+\mathbf{Q}_1} & w_{23}  \\
w_{31} & w_{32} & \epsilon_{\mathbf{k}-\mathbf{Q}_1}    \\
\end{array} \right) ,
\label{Hk1}
\end{equation}
where the off-diagonal entries of the Hamiltonian are
\begin{eqnarray}
w_{12}&=w_{21}^*=W_0\left(\cos\left(k_x+{Q}/{2}\right)-\cos k_y\right),\\
w_{13}&=w_{31}^*=W_0\left(\cos\left(k_x-{Q}/{2}\right)-\cos k_y\right),\\
w_{23}&=w_{32}^*=W_0\left(\cos\left(k_x+3Q/2\right)-\cos k_y\right)
\end{eqnarray}
The above Hamiltonian can be diagonalized: $H(\mathbf{k})=\sum\limits_n E_{\mathbf{k},n}a^\dagger_{\mathbf{k},n}a_{\mathbf{k},n}$, where $a_{\mathbf{k},n}$ are the quasiparticle operators, which can be represented in terms of the fermion operator as $c_{\mathbf{k},\sigma}=\sum\limits_n U_{\mathbf{k},1,n}a_{\mathbf{k,n}}$, $c_{\mathbf{k}+\mathbf{Q},\sigma}=\sum\limits_n U_{\mathbf{k},2,n}a_{\mathbf{k,n}}$, $c_{\mathbf{k}-\mathbf{Q},\sigma}=\sum\limits_n U_{\mathbf{k},3,n}a_{\mathbf{k,n}}$, where $U_{\mathbf{k},i,j}$ are elements of the unitary transformation $U_{\mathbf{k}}$ which diagonalizes $H(\mathbf{k})$. As shown below, the factors $U_{\mathbf{k}i,j}$ are important for the evaluation of the single-particle spectral function, which is measurable in angle resolved photoemission experiments \cite{Chakravarty-ARPES}.

In cuprates, the hole doping is conventionally counted from half-filling, i.e., one electron per Cu atom. If $g$ denotes the fraction of an occupied number of states in the Brillouin zone, then the doping is $p=1-2g$. The fraction $g$ is calculated as
$g=1/\mathcal{S}\sum\limits_{n,{\mathbf{k}\in RBZ}}{f(E_{\mathbf{k},n})}$,
where $f(E_n(\mathbf{k}))$
is the Fermi distribution function and  $E_n(\mathbf{k})$ is the quasiparticle energy dispersion for the $n^{th}$ energy band, measured with respect to the chemical potential (the chemical potential is evaluated for each doping value using the equation for $g$), and $\mathcal{S}$ is the area of the system.

The quasiparticle Fermi surface for the unidirectional CDW state given by $\sum\limits_n \delta(\omega-E_{\mathbf{k},n})$ is shown in Fig.~\ref{Fig_FS_ARPES_1}, top panel, at $\omega=0$ for a typical hole doping $p\sim 0.16$ close to optimal doping.
With a finite CDW order parameter, the Fermi surface reconstructs from a large hole-like surface in the overdoped regime ($p > p^{*}$) to isolated  hole pockets centered around $(\pi/3,\pi)$ in the underdoped regime ($p < p^*$). To show that the emergence of the hole pockets is a robust consequence of unidirectional CDW, in Fig.~\ref{Fig_FS_ARPES_1} bottom panel we plot the Fermi surface for a $4\times4$ reconstruction of the Brillouin zone, with a CDW order parameter $\textbf{Q}=(0.25,0)r.l.u$. In both cases $\textbf{Q}=(\pi/3,0)r.l.u$ and $(0.25,0)r.l.u$ the unidirectional CDW reconstructs the Fermi surface into hole pockets, (in contrast, the bidirectional CDW has electron pockets, see Fig.~\ref{Fig_Coex_FS}).

\section{Hall number in unidirectional CDW}
As noted in the introduction, the zero-temperature Hall number provides information about the volume enclosed by the Fermi surface. Further, the sign of the Hall number reveals the nature of dominant carriers (electrons or holes). A drastic drop in $n_H$ below optimal doping indicates a drastic reconstruction of a large Fermi surface enclosing a volume corresponding
to a density $n_c \simeq 1 + p$ of holes at large doping, to small pockets with a volume corresponding to hole-density $p$ in the underdoped regime.

In linear response theory the conductivities $\sigma_{xx}, \sigma_{xy}$ are computed using the Kubo formulae, which in the limit of long scattering time $\tau$, and for $\textbf{q} \rightarrow 0, \omega \rightarrow 0$,  reduce to the following Boltzmann formulae (see Appendix) 
\begin{eqnarray}
\sigma_{xx} = e^2\sum_n\int{\tau_\mathbf{k} (v_n^x)^2 \left(-\frac{\partial f[E_n(\mathbf{k})]}{\partial E_n(\mathbf{k})}\right)d^2\mathbf{k}}
\label{sxx}
\end{eqnarray}
\begin{eqnarray}
\sigma_{xy} = \frac{e^3}{\hbar}\sum\limits_n\int{\tau^2_\mathbf{k} \left(-\frac{\partial f[E_n(\mathbf{k})]}{\partial E_n(\mathbf{k})}\right)v_n^x \left(v_n^y v_n^{xy} - v_n^x v_n^{yy}\right)d^2\mathbf{k}}\nonumber \\
\label{sxy}
\end{eqnarray}
where $n$ is the band index, $v^x_n$ is the semi-classical quasi-particle velocity $v^x_n=\frac{1}{\hbar}\frac{\partial E_n(\mathbf{k})}{\partial k_x}$ and $v_n^{xy}=\frac{\partial v^y_n}{\partial k_y}$. The factor $\tau_{\mathbf{k}}$ is the phenomenological scattering time. 
Since in our calculations the energy dispersion $E_n^k$ depends on the value of the CDW order parameter, which in turn is a function of hole doping, it follows that the change in the effective mass (hidden in the definitions of the  derivatives of $E_n(\mathbf{k})$) with hole doping is accounted for in our formalism. 
The Hall coefficient is given by $R_H=\sigma_{xy}/\sigma_{xx}\sigma_{yy}$. We  compute the Hall number in the relaxation time approximation, ignoring intra-band scattering effects. We also assume that the CDW quasiparticles have a constant scattering time. Strictly speaking, the scattering time may vary along the Fermi surface, but this additional complication does not qualitatively alter our results in the unidirectional state. 

In Fig.~\ref{Fig_Hall_number} we plot the Hall number $(n_{H}=R_H^{-1})$ obtained from our calculations for the doping range $0.16<p<0.20$. Below $p^{*}\sim 0.19$, the Fermi surface topology is that of a unidirectional CDW state, namely, hole pockets. For doping dependence of the CDW strength ($W(p)$), we examine two different phenomenological functional forms ($W\sim W_0 (p^*-p)$) and $W(p)\sim W_0(p^*-p)^{1/2}$ for illustrative purposes \cite{Tewari}, making a simple assumption that $W$ is a smoothly varying function of the hole doping. We observe a rapid drop in the Hall number below $p^{*}$. The magnitude of the drop in $n_H$ below $p^*$ depends on the choice of various parameters in particular on the value of $W_0$. For instance a larger value of $W_0$ shrinks the size of hole pockets and leads to larger suppression of $n_H$ at $p\sim p^*$. Above $p^{*}$, the Hall number crosses to $n_H \sim 1+p$ as expected from conventional Fermi liquid theory. In the weak-field regime, the width of the Fermi surface reconstruction depends on the magnitude of the CDW gap. In Fig.~\ref{Fig_Coex_Rh} inset we show that a weak bidirectional component coexisting with dominant unidirectional order still results in the suppression of Hall number without change of sign near optimal doping.

In Fig.~\ref{Fig_rh_rhoxx_sxy} we plot the Hall resistance ($R_H$), longitudinal resistivity ($\rho_{xx}$), and Hall conductivity ($\sigma_{xy}$) as a function of temperature for various doping values in the actual experimental units, as predicted by a uni-directional CDW order. We use a standard mean field dependence of the CDW order parameter ($W\sim\sqrt{1-T/T_{CDW}})$~\cite{Hussey}. The order of magnitude of the above quantities, and the behavior w.r.t. temperature and doping match with the corresponding experimentally observed quantities in Ref~\onlinecite{Badoux:2016}.

\section{Coexisting unidirectional and bidirectional orders}
Recent NMR \cite{Wu:2015}, high-field x-ray scattering \cite{Gerber,Chang,Jang}, and sound velocity \cite{Lali} measurements at moderately underdoped regime reveal the coexistence of a bidirectional order parameter along with the unidirectional CDW at high magnetic fields and low temperatures. In our calculations, for doping below $p\sim 0.16$, we assume the presence of unidirectional and bidirectional CDWs of equal magnitude, which essentially is a bidirectional CDW with anisopropic strengths of the order parameter. 

The bidirectional mean-field CDW state can be represented by the Hamiltonian,
\begin{eqnarray}
H_{CDW}^{bi} &= \sum\limits_{\mathbf{r},\mathbf{a},\sigma} [V_{\mathbf{a}} \left(e^{i\mathbf{Q}_1\cdot(\mathbf{r}+\mathbf{a}/2)} + e^{i\mathbf{Q}_2\cdot(\mathbf{r}+\mathbf{a}/2)}\right)c^{\dagger}_{\mathbf{r}+\mathbf{a},\sigma}c_{\mathbf{r},\sigma}\nonumber \\ & + h.c],
\label{BDW_bi},
\end{eqnarray}
where $V_{\pm\hat{x}} = -V_{\pm\hat{y}} = V_0/2$, $V_0$ is the bare magnitude of the bidirectional order parameter, $\mathbf{Q}_1=(q,0)$, and $\mathbf{Q}_2=(0,q)$. Choosing $q=(1/3)r.l.u$ now gives us a nine-component reconstruction of the Brillouin zone. In the coexistence phase, the total CDW Hamiltonian is given by
\begin{eqnarray}
H_{CDW} = H_{CDW}^{uni} + H_{CDW}^{bi}
\label{EQ_HCDW_coex}
\end{eqnarray}
In the above equation, the unidirectional CDW Hamiltonian is now also represented by a nine-component reconstructed BZ, however with a different bare-magnitude ($W_0$) and scattering vector present only in one direction. Details of the Hamiltonian for this phase have been provided in the Appendix.

The addition of a bidirectional order parameter to the unidirectional CDW changes the Fermi surface from isolated hole pockets in the first Brillouin zone (unidirectional CDW) to electron pockets centered at $(\pi/3,\pi/3)$ in the coexistence phase. The electron pockets in the coexisting phase are visible in Fig.~\ref{Fig_Coex_FS}. In angle-resolved photoemission (ARPES) experiments the quasiparticle spectral function $A(\omega,\mathbf{k})$ is mapped on the Brillouin zone. The spectral function is given by
\begin{eqnarray}
A(\omega,\mathbf{k}) = -\frac{1}{\pi}\mbox{Im } G_{\mbox{ret}} (\omega,\mathbf{k}),
\end{eqnarray}
where $G_{\mbox{ret}} (\omega,\mathbf{k})$ is the retarded Green's function for the Hamiltonian. In terms of the unitary transformation $U_{\mathbf{k}}$ discussed previously, $A(\omega,\mathbf{k})$ becomes
\begin{eqnarray}
A(\omega,\mathbf{k}) = \sum\limits_n{U^2_{\mathbf{k},1,n}\delta(\omega-E_{\mathbf{k},n})}
\end{eqnarray}
where the band index $n$ counts the total number of bands in the reconstructed Hamlitonian,
and $U_{\mathbf{k},1,n}$ are the coherence factors given by entries of the transformation $U_\mathbf{k}$. Due to strong momentum dependence of the spectral weights, the ARPES spectral function of the coexisting phase consists of Fermi arcs. Fig.~\ref{Fig_Coex_FS} shows the quasiparticle energy and angle resolved photoemission spectrum with electron pockets centered at $(\pi/3,\pi/3)$ for CDW state with coexisting unidirectional  and bidirectional  order parameters. The curves in the upper plot of Fig.~\ref{Fig_Coex_FS} represent the energy bands. The intersection of the bands with chemical potential depicts the electron pocket.

\begin{figure}
\includegraphics[scale=0.24]{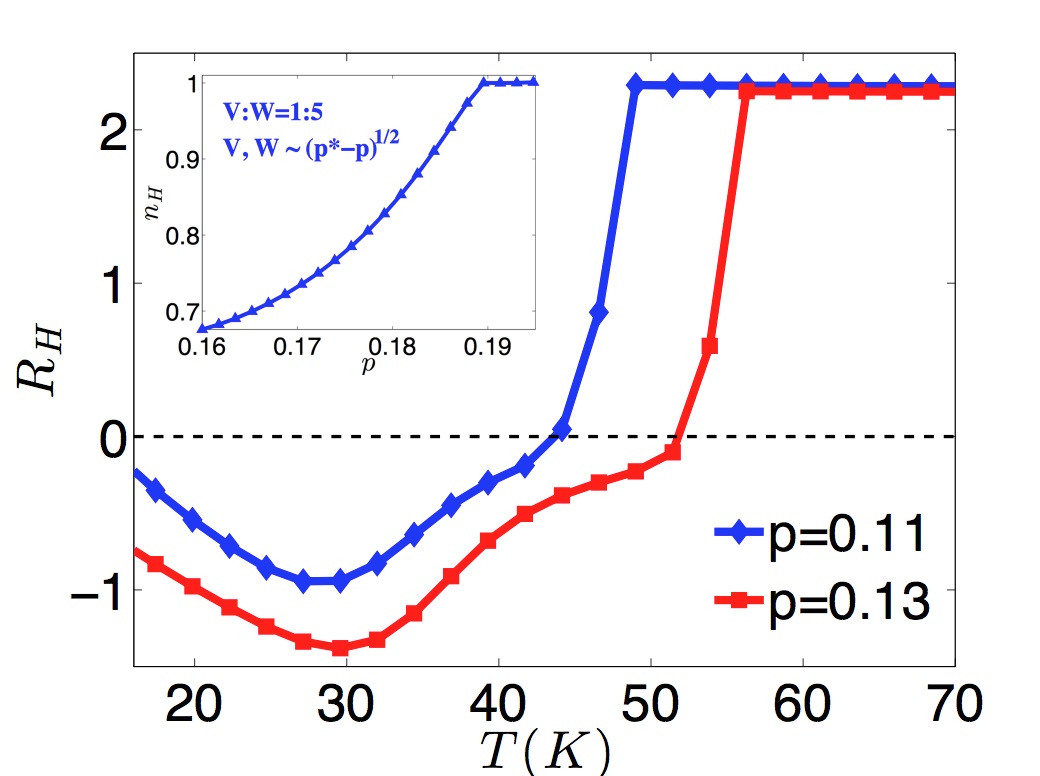}
\caption{(color online) Hall resistance $R_H$ (in arbitrary units) from the coexisting unidirectional and bidirectional CDW state plotted as a function of temperature for two different doping values $p=0.11$, $p=0.13$. The change of sign of $R_H$ from positive to negative values as the temperature is lowered is due to the bidirectional order. The bidirectional phenomenological mean field order parameter is taken as  $V(p,T)\sim V(p) \sqrt{1-T/T_{CDW}(p)}$, and $V(p)\sim V_0\sqrt{p-p_c}$, where the bare magnitude $V_0$ was chosen to be $0.22t_1$, $p_c$ was chosen to be $p_c=0.085$. The strengths of unidirectional and bidirectional order were chosen to be equal. \textit{Inset:} Hall number $n_H$ (normalized w.r.t the Hall number at $p=0.19$) from the coexisting unidirectional and bidirectional CDW order plotted as a function of hole doping. The ratio of bidirectional ($V$) and unidirectional order parameters ($W$) was taken to be $1:5$ with a square root doping dependence below $p^*=0.19$. The suppression of the Hall number below $p^*$ is still observed even after adding a small bidirectional order parameter.}
\label{Fig_Coex_Rh}
\end{figure}

For coexisting order parameters, since the bidirectional CDW introduces an electron pocket to the Fermi surface, for calculation of the Hall resistance we retain a small momentum dependence of the scattering time \cite{Seo:2014} to enhance the contribution from the electron pockets. We plot the Hall resistance $(R_H=1/n_H)$ as a function of temperature in Fig.~\ref{Fig_Coex_Rh}. Since a bidirectional CDW has an electron pocket, the zero temperature Hall coefficient reduces to negative values, consistent with the experiments \cite{LeBoeuf:2007,Chang:2010,Laliberte:2011}. In the inset of Fig.~\ref{Fig_Coex_Rh} we plot the Hall number below $p^*=0.19$ adding a small bidirectional component to the uniaxial order parameter near optimal doping. The suppression of the Hall number below $p^*$ is still observed even after adding a week bidirectional component. 

Therefore, we have shown that the high field unidirectional CDW order has hole
pockets, in contrast to its bidirectional counterpart which has both electron and hole pockets. The existence of the hole pockets leads to the suppression of Hall number but no change of
sign near optimal doping. The addition of a weak bidirectional component to the order
parameter, as in the inset of Fig.~\ref{Fig_Coex_Rh}, does not change the behavior of $n_H$ near optimal
doping. Only with a sufficiently strong bidirectional component of the order parameter the Hall
coefficient changes sign, as in our intermediate doping range $0.16 > p > 0.08$. Our calculations therefore show that the complex evolution of high field Hall number in the entire pseudogap phase can be explained in terms of a coexistence of unidirectional and bidirectional charge order, with weak bidirectional component near optimal doping.   
\section{Discussion}
Several comments are in order: First, we have assumed in this paper that the dominant instability at high magnetic fields ($H \sim H_{c2}$) and low temperatures near optimal doping in hole-doped cuprate superconductors is to a unidirectional CDW state while the amplitude of a bidirectional CDW is small. In contrast to bidirectional CDW, we have shown that the Fermi surface in the unidirectional CDW displays hole pockets. The emergence of the hole pockets, which is a result of the first Fermi surface reconstruction with decreasing hole doping at the quantum critical point near optimal doping \cite{Tallon}, does not change the sign of the Hall coefficient but results in a rapid suppression of carrier concentration and zero temperature Hall number with decreasing doping below $p \sim 0.19$. Although this is consistent with experiments on high field Hall effect near optimal doping \cite{Bala1,Bala2,Bala3,Badoux:2016,Laliberte:2016,Collignon:2016}, the identification of the quantum critical point near optimal doping with unidirectional charge order is at present not supported by experiments \cite{Badoux2}. No such charge order has so far been found near optimal doping, directly or indirectly \cite{Damascelli,Wu:2015,Wu:2011,Wu:2013,Chang:2012,Ghiringhelli:2012,Comin,Neto,Tabis,Croft,Gerber,Chang,Jang,Blackburn,Blanco,Lali}. On the other hand, in contrast to other scenarios attempting to explain the suppression of Hall number near optimal doping \cite{Tewari,Eberlein,Maharaj,Storey,Yang,Kaul,Qi,Sachdev1,Chatterjee}, both unidirectional and bidirectional charge order have been found in the nearby range of hole doping  \cite{Damascelli,Wu:2011,Wu:2013,Chang:2012,Ghiringhelli:2012,Comin,Neto,Tabis,Croft,Gerber,Chang,Jang,Blackburn,Blanco,Lali} and a coexistence phase has been established for moderate doping $0.8\lesssim p \lesssim 0.16$ \cite{Gerber,Chang,Jang,Lali}. We have used these latter experiments to add a bidirectional component to the order parameter for intermediate range of hole doping leading to a second reconstruction of the Fermi surface to an electron pocket. The corresponding zero temperature Hall number reduces to negative values in the coexisting phase. 
The consistency of $n_H$  with CDW in the underdoped regime, and the lack of experimental observation of CDW up to $p^*$ at this time, may actually mean that more experiments are needed at very high magnetic fields ($H \sim H_{c2}$ ) close to the pseudogap critical doping $p^*$.

 Second, with respect to the high field Hall measurements in the cuprates, the older experiments \cite{Bala1,Bala2,Bala3} found a sharp decrease of $n_H$ with decreasing hole doping below optimal doping $p^*$ just as the more recent ones \cite{Badoux:2016,Laliberte:2016,Collignon:2016}. However, Hall number in the older experiments also dropped for $p > p^*$, producing a peak like structure of $n_H$ versus $p$ at optimal doping $p^*$. The more recent studies \cite{Badoux:2016,Laliberte:2016,Collignon:2016}, performed at somewhat higher temperatures because of higher values of the upper critical fields, on the other hand, find that $n_H$ saturates with $p$ for $p > p^*$. While it has been argued \cite{Collignon:2016} that the presence of electron-like carriers co-existing with hole-like carriers in a narrow range of doping close to optimal doping may be responsible for non-universal behavior of $n_H$ at $p=p^*$, possibly even resulting in a peak, we have not attempted to address this issue here.

Third, our theoretical framework is that of effective Hartree-Fock mean field Hamiltonians that capture the broken symmetries associated with the charge orders. By definition, therefore, we ignore the fluctuation effects that must be present very close to the quantum critical point at $p=p^*$ where the large hole-like Fermi surface at $p>p^*$ with carrier concentration $n\sim 1+p$ is assumed to reconstruct into small hole pockets with $n \sim p$ via a transition into a unidirectional CDW state, and also near the Fermi surface reconstruction at a lower value of $p$ where the system enters the co-existence phase. Our choice of the dependence of the order parameter amplitudes on doping, $W \sim W_0(p^*-p)^{1/2}$ and $W\sim W_0(p^*-p)$, are thus phenomenological and \textit{ad hoc}, and cannot be valid very close to the critical doping $p^*$. In particular, our calculations cannot be valid within a narrow critical fluctuation region $\delta p \sim (\frac{W_0}{\epsilon_F})^2$ where $\epsilon_F$ is the effective Fermi energy of the holes around the critical doping $p^*$.

Finally, it is now well established that in the cuprates the two kinds of charge order, although coexisting at high fields and low temperatures in a wide range of hole doping, are somewhat different in nature even aside from the difference in the symmetries dictated by their modulation wave vectors. While the unidirectional order, which emerges only above high magnetic fields ($H \sim H_{c2}$) and low temperatures $T \lesssim T_c$, is long range ordered and three dimensional (3D), the bidirectional order exists even for small or zero magnetic fields and at temperatures $T \lesssim 150 K$ somewhere between superconducting $T_c$ and the pseudogap temperature scale $T^*$. Further, the bidirectional order is essentially two dimensional (2D) and with significantly smaller in-plane correlation lengths than the unidirectional order. Interestingly, however, the magnitudes of their modulation wave vectors in the in-plane directions are identical suggesting the same intrinsic correlations in the electronic wave functions responsible for both types of order. Despite the similarities in the in-plane wave vectors and a wide range of co-existence in the pseudogap regime, at present the relationship between the two charge orders is unclear at best and we have ignored any such issues in this paper.

\section{Conclusion}
To summarize, we have shown that the onset of a  low-temperature high-field unidirectional incommensurate CDW with or without a weak bidirectional component in copper oxide superconductors may help explain the rapid drop in Hall number below optimal doping as seen in recent experiments. The single-particle spectral function in the high-field unidirectional CDW displays hole pockets. The emergence of the hole pockets is a result of Fermi surface reconstruction at the quantum critical point, resulting in a a rapid suppression of the Hall number with decreasing hole doping. Adding a bidirectional component of approximately same magnitude to the order parameter at lower doping introduces an electron pocket in the Fermi surface. The corresponding zero temperature Hall number reduces to negative values in the coexisting phase. Our calculations explain the salient features of the recent high field Hall effect experiments in the cuprate superconductors in terms of unidirectional and bidirectional charge orders both of which have been unambiguously observed in the pseudogap phase  at the relevant ranges of magnetic field and temperature.

The primary result of this work is the explanation of the rapid suppression of the Hall number near optimal doping (while staying positive at low temperatures), indicating a possible Fermi surface reconstruction from large hole-like Fermi surface to small hole pockets below optimal doping, in terms of a uni-directional CDW state. We have also included a discussion of the behavior of the Hall coefficient in the moderately underdoped regime (where it changes sign at low temperatures), which was discussed before in Ref.~\onlinecite{Seo:2014}, for the sake of completeness.\\

\textit{Acknowledgments}
AT acknowledges CSIR (India) funding under the project grant no: 03 (1373)/16/EMR-II. ST acknowledges support from ARO Grant No: (W911NF-16-1-0182).\\


\appendix
\section{Calculation of conductivities}
In linear response theory the conductivities $\sigma_{xx}, \sigma_{xy}$ are computed using the following Kubo formulae
\begin{equation}
\sigma_{xx}=\frac{1}{\pi} \rm{Im} \Pi(i\omega_n\rightarrow \omega + i\delta, {q}=0, T),
\label{Kubo1}
\end{equation}
and
\begin{equation}
\sigma_{xy}=\rm{lim}_{{q}\rightarrow 0}\frac{B}{\omega {q}}\rm{Re}\tilde{\Pi}(i\omega_n\rightarrow \omega + i\delta, q\hat{y},T),
\label{Kubo2}
\end{equation}
which are valid in the presence of weak electric and magnetic fields $\textbf{E}=E_0\hat{x}\cos(\omega t), \textbf{B}=q\hat{z}A_0\sin(qy)$. Here, the correlation functions are given by,
\begin{equation}
\Pi(i\omega_n, \textbf{q}, T)= \int_0^{\beta}d\tau e^{i\omega_n\tau}\langle T_{\tau}j_x(\textbf{q},\tau)j_x(\textbf{q},\tau)\rangle,
\end{equation} and
\begin{eqnarray}
\tilde{\Pi}&&(i\omega_n, \textbf{q},T)\nonumber\\
&&=\int_0^{\beta}d\tau d\tau'e^{i\omega_n\tau}\langle T_{\tau}j_y(\textbf{q},\tau)j_x(0,0)j_y(-\textbf{q},\tau')\rangle,
\end{eqnarray}
with $j_x,j_y$ appropriately defined current operators in the CDW state. In the limit of long scattering time, and for
$\textbf{q} \rightarrow 0, \omega \rightarrow 0$, the linear response expressions for the conductivities reduce to those in Boltzmann description provided in the main text of the paper.

\section{Hamiltonian for coexisting orders}

Here we provide the Hamiltonian for the coexisting unidirectional and bidirectional CDW phase.
The Hamiltonian can most easily be written by coupling wave vector $\textbf{k}$, confined to a properly defined first Brillouin zone (FBZ), with wave vectors translated by the CDW wave vector $\textbf{Q}$, i.e., $\textbf{k} \rightarrow \textbf{k}+ n_x Q_x \hat{x}+n_y Q_y \hat{y}$ where $n_x,n_y$ are integers denoting translations in the two-dimensional reciprocal space. Strictly speaking, for incommensurate systems this procedure results in an infinite dimensional Hamiltonian matrix and infinite number of bands. However, for the incommensuration $\delta \ll Q$, we can approximate the relevant energy eigenvalues by partitioning the unfolded BZ and defining energy bands over each BZ sector. The mean-field Hamiltonian for the bidirectional phase (with bare-magnitude $V_0$) then becomes
\begin{widetext}
\begin{equation}
H(\mathbf{k})_{CDW}^{bi}=\left( \begin{array}{ccccccccc}
\epsilon_{\mathbf{k}} & v_{12} & v_{13} &  v_{14} & 0 & 0 & v_{17} & 0 & 0 \\
v_{21} & \epsilon_{\mathbf{k}+\mathbf{Q}_1} & v_{23} & 0 & v_{25} & 0 & 0 & v_{28} & 0  \\
v_{31} & v_{32} & \epsilon_{\mathbf{k}-\mathbf{Q}_1} & 0 & 0 & v_{36} & 0 & 0 & v_{39}   \\
v_{41} & 0 & 0 & \epsilon_{\mathbf{k}+\mathbf{Q}_2} & v_{45} & v_{46} & v_{47} & 0 & 0   \\
0 & v_{52} & 0 & v_{54} & \epsilon_{\mathbf{k}+\mathbf{Q}_1+\mathbf{Q}_2} & v_{56} & 0 & v_{58} & 0   \\
0 & 0 & v_{63} & v_{64} & v_{65} & \epsilon_{\mathbf{k}-\mathbf{Q}_1+\mathbf{Q}_2} & 0 & 0 & v_{69}   \\
v_{71} & 0 & 0 & v_{74} & 0 & 0 & \epsilon_{\mathbf{k}-\mathbf{Q}_2} & v_{78} & v_{79}   \\
0 & v_{82} & 0 & 0 & v_{85} & 0 & v_{87} & \epsilon_{\mathbf{k}+\mathbf{Q}_1-\mathbf{Q}_2} & v_{89}  \\
0 & 0 & v_{93} & 0 & 0 & v_{96} & v_{97} & v_{98} & \epsilon_{\mathbf{k}-\mathbf{Q}_1-\mathbf{Q}_2}  \\
\end{array} \right)
\end{equation}
\end{widetext}
The non zero elements of the above 9 component Hamiltonian are specifically given by
\begin{widetext}
\begin{eqnarray*}
v_{12}&=V_0\left(\cos\left(k_x+{\pi}/{3}\right)-\cos k_y\right),
v_{13}&=V_0\left(\cos\left(k_x-{\pi}/{3}\right)-\cos k_y\right),\\
v_{14}&=V_0\left(\cos k_x-\cos\left(k_y+{\pi}/{3}\right)\right),
v_{17}&=V_0\left(\cos k_x-\cos\left(k_y-{\pi}/{3}\right)\right),\\
v_{23}&=V_0\left(\cos\left(k_x+\pi\right)-\cos k_y\right),
v_{25}&=V_0\left(\cos\left(k_x+{2\pi}/{3}\right)-\cos \left(k_y+{\pi}/{3}\right)\right),\\
v_{28}&=V_0\left(\cos\left(k_x+{2\pi}/{3}\right)-\cos \left(k_y-{\pi}/{3}\right)\right),
v_{36}&=V_0\left(\cos\left(k_x-{2\pi}/{3}\right)-\cos \left(k_y+{\pi}/{3}\right)\right),\\
v_{39}&=V_0\left(\cos\left(k_x-{2\pi}/{3}\right)-\cos \left(k_y-{\pi}/{3}\right)\right),
v_{45}&=V_0\left(\cos\left(k_x+{\pi}/{3}\right)-\cos \left(k_y+{2\pi}/{3}\right)\right),\\
v_{46}&=V_0\left(\cos\left(k_x-{\pi}/{3}\right)-\cos \left(k_y+{2\pi}/{3}\right)\right),
v_{47}&=V_0\left(\cos k_x-\cos \left(k_y+\pi\right)\right),\\
v_{56}&=V_0\left(\cos\left(k_x+\pi\right)-\cos \left(k_y+{2\pi}/{3}\right)\right),
v_{58}&=V_0\left(\cos\left(k_x+{2\pi}/{3}\right)-\cos (k_y+\pi)\right),\\
v_{69}&=V_0\left(\cos\left(k_x-{2\pi}/{3}\right)-\cos (k_y+\pi)\right),
v_{78}&=V_0\left(\cos\left(k_x+{\pi}/{3}\right)-\cos \left(k_y-{2\pi}/{3}\right)\right),\\
v_{79}&=V_0\left(\cos\left(k_x-{\pi}/{3}\right)-\cos \left(k_y-{2\pi}/{3}\right)\right),
v_{89}&=V_0\left(\cos\left(k_x+\pi\right)-\cos \left(k_y-{2\pi}/{3}\right)\right).\\
\end{eqnarray*}
\end{widetext}
With the above BZ reconstruction, the Hamiltonian for the unidirectional CDW phase (with bare-magnitude $W_0$) is
\begin{widetext}
\begin{equation}
H(\mathbf{k})_{CDW}^{uni}=\left( \begin{array}{ccccccccc}
\epsilon_{\mathbf{k}} & w_{12} & w_{13} &  0 & 0 & 0 & 0 & 0 & 0 \\
w_{21} & \epsilon_{\mathbf{k}+\mathbf{Q}_1} & w_{23} & 0 & 0 & 0 & 0 & 0 & 0  \\
w_{31} & w_{32} & \epsilon_{\mathbf{k}-\mathbf{Q}_1} & 0 & 0 & 0 & 0 & 0 & 0   \\
0 & 0 & 0 & \epsilon_{\mathbf{k}+\mathbf{Q}_2} & w_{45} & w_{46} & 0 & 0 & 0   \\
0 & 0 & 0 & w_{54} & \epsilon_{\mathbf{k}+\mathbf{Q}_1+\mathbf{Q}_2} & w_{56} & 0 & 0 & 0   \\
0 & 0 & 0 & w_{64} & w_{65} & \epsilon_{\mathbf{k}-\mathbf{Q}_1+\mathbf{Q}_2} & 0 & 0 & 0   \\
0 & 0 & 0 & 0 & 0 & 0 & \epsilon_{\mathbf{k}-\mathbf{Q}_2} & w_{78} & w_{79}   \\
0 & 0 & 0 & 0 & 0 & 0 & w_{87} & \epsilon_{\mathbf{k}+\mathbf{Q}_1-\mathbf{Q}_2} & w_{89}  \\
0 & 0 & 0 & 0 & 0 & 0 & w_{97} & w_{98} & \epsilon_{\mathbf{k}-\mathbf{Q}_1-\mathbf{Q}_2}  \\
\end{array} \right)
\end{equation}
\end{widetext}
where the elements can be calculated similar to the bidirectional case, albeit with a different magnitude of order parameter ($W_0$). In the coexistence phase, the total CDW Hamiltonian is given by
\begin{eqnarray}
H_{CDW} = H_{CDW}^{uni} + H_{CDW}^{bi}
\end{eqnarray}
as given in Eq. 16 of the main text.

\end{document}